# Modelling species distributions using remote sensing predictors: Comparing Dynamic Habitat Index and LULC

Maïri Souza Oliveira[a*], Clémentine Préau[a], Samuel Alleaume[a], Maxime Lenormand [a*], Sandra Luque[a]

[a]INRAE, National Research Institute on Agriculture, Food & the Environment, TETIS Unit, Maison de la télédétection, 34090 Montpellier, France

*Corresponding authors: mairi.souza-oliveira@inrae.fr & maxime.lenormand@inrae.fr

## Abstract

This study compares the predictive capacity of the Dynamic Habitat Index (DHI)—a remote sensing (RS)-based measure of habitat productivity and variability— against traditional land-use/land-cover (LULC) metrics in species distribution modelling (SDM) applications. RS and LULC-based SDMs were built using distribution data for eleven bird, amphibian, and mammal species in Île-de-France. Predictor variables were derived from Sentinel-2 RS data and LULC classifications, with the latter incorporating Euclidean distance to habitat types. Ensemble SDMs were built using nine algorithms and evaluated with the Continuous Boyce Index (CBI) and a calibrated AUC. Habitat suitability scores and their binary transformations were assessed using niche overlap indices (Schoener, Warren, and Spearman rank correlation coefficient). Both RS and LULC approaches exhibited similar predictive accuracy overall. After binarisation however, the resulting niche maps diverged significantly. While LULC-based models exhibited spatial constraints (habitat suitability decreased as distance from recorded occurrences increased), RS-based models, which used continuous data, were not affected by geographic bias or distance effects. These results underscore the need to account for spatial biases in LULC-based SDMs. The DHI may offer a more spatially neutral alternative, making it a promising predictor for modelling species niches at regional scales.

## Graphical abstract

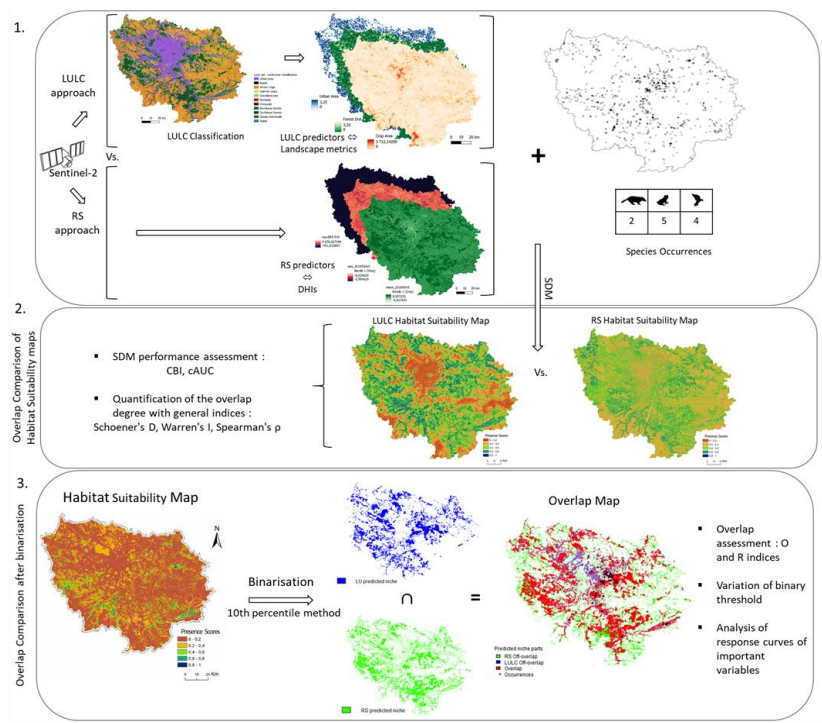

# 1. Introduction

Species Distribution Modelling (SDM), also known as Ecological Niche Modelling (ENM), has become one of the most widely used tools for studying and forecasting species distributions and is extensively applied in biogeography, ecology, and conservation (Zurell et al., 2020). The aim of SDM approaches is to predict species distribution based on the statistical correlation between their occurrences and environmental variables that are presumed to influence their distribution (Guisan and Zimmermann, 2000). The selection of predictor variables depends on several factors, including the scale and resolution of the study area and the accuracy of the occurrence data. Many SDM studies incorporate environmental variables derived from spatial interpolation methods, such as bioclimatic variables (Barbet-Massin and Jetz, 2015; Booth, 2018), which are commonly used at continental and global scales (Pearson, 2007; Thuiller et al., 2004). At local and regional scales however, studies often rely on variables based on categorised data such as topography and land-use/land-cover (LULC) data (Cord et al., 2014; Luoto et al., 2007).

Predictor variables should be carefully selected for use in species distribution models (SDMs) to ensure that they are ecologically meaningful and appropriate to the specific study context. The availability of climate and LULC data—interpolated and categorised, respectively—has facilitated their integration into SDMs and the forecasting of species range shifts under global change scenarios (e.g. Bellard et al., 2012; Guo et al., 2019). Commonly used LULC predictors include the proportion of land cover classes (Chen and Leites, 2020; Marshall et al., 2021), distances to specific LULC types (Bellón et al., 2022; Dubos et al., 2022), and metrics characterising landscape composition and configuration (Coops et al., 2016; Radeloff et al., 2019).

In addition to the assumptions and uncertainties inherent to SDMs (Austin, 2007; Dubos et al., 2022; Elith et al., 2006; Zimmermann et al., 2009), LULC predictors are also constrained to varying degrees by the quality and resolution of available data for the species and geographic region under study (Chen & Leites, 2020). Since habitat suitability and resource availability cannot be exhaustively captured by LULC predictors, small or heterogeneous habitat patches typically critical for landscape structure and land-use management (Randin et al., 2020) may be omitted. Such limitations may increase the risk of inferring spurious causal relationships, a common limitation in species distribution modelling studies (Currie et al., 2020; Fourcade et al., 2018), and may reduce model transferability across regions or contexts (Petitpierre et al., 2016). When incorporating LULC variables into local-scale SDMs, it is important to recognise that they are not always well-suited as ecological niche descriptors; they are categorical variables indirectly derived from spectral data which may introduce additional sources of uncertainty. It can be difficult to determine the appropriate number and types of LULC classes to include. These classes, typically derived from supervised classification, are often created using diverse data sources that may introduce errors, uncertainties, and imprecision (Cord et al., 2014). These data are often static and do not capture temporal changes in land use (Hansen and Loveland, 2012; Pellissier et al., 2013). Consequently, SDMs based exclusively on LULC data may fail to account for the potential effects of short- and long-term environmental changes on species occurrence (Coops and Wulder, 2019; Leitão and Santos, 2019). Researchers therefore advocate the use of more spatially explicit and ecologically meaningful predictors (Arenas-Castro et al., 2018; Gardner et al., 2019; Mod et al., 2016; Scherrer and Guisan, 2019), especially when using remote sensing (RS) data (He et al., 2015; Leitão and Santos, 2019; Regos et al., 2022).

The development of RS products (e.g. Alleaume et al., 2018; Rocchini et al., 2016) has shown great potential to support biodiversity monitoring and conservation efforts (Randin et al., 2020; Rose et al., 2015). Access to RS data from multitemporal series at very high spatial resolution (HSR) – provided by satellite missions with frequent revisit times and extensive spatial coverage, such as Copernicus Sentinel-2 (S2) – enables the inclusion of spatially explicit variables, directly or indirectly causally related to species distribution (Alcaraz-Segura et al., 2017; Arenas-Castro et al., 2018; He et al., 2015; Pettorelli et al., 2018). Using spectral data to derive continuous radiometric indices of vegetation, soil, and land cover (Wilson et al., 2013). These variables include information on seasonality, phenology and productivity (Luque et al., 2018; Pettorelli et al., 2011). Of these variables, the Normalised Difference Vegetation Index (NDVI) is the most frequently used in SDM studies (Clauzel and Godet, 2020; Tarabon et al., 2019).

Ecosystem structure and functioning, approximated by these radiometric indices, are sensitive to environmental disturbances and annual and interannual environmental variations, and thus temporal data series are required to monitor such changes accordingly (Fisher et al., 2010; Regos et al., 2022). Indices that approximate ecosystem dynamics are therefore crucial to monitoring efforts (Arenas-Castro et al., 2018; Mod et al., 2016; Scherrer and Guisan, 2019). Continuous series of multitemporal satellite images provide predictive variables that capture ecosystem states and environmental changes at ecologically relevant time scales (Randin et al., 2020). This is exemplified by the Dynamic Habitat Index (DHI) which summarises annual variations of radiometric indices by calculating 3 components: (i) DHI_cum, representing the cumulative radiometric index, (ii) DHI_min, representing the minimum value, and (iii) DHI_cv, representing the coefficient of variation of values over the year (Berry et al., 2007; Coops et al., 2008). The DHI was initially developed to be primarily applied to the NDVI to estimate annual variations in the primary productivity of habitat vegetation (Bonthoux et al., 2018; Coops et al., 2009a, 2009b). This index has shown correlations with amphibian, bird, and mammal species richness on a global scale (Hobi et al., 2017; Radeloff et al., 2019). By mapping areas with high habitat suitability and connectivity, the DHI can delineate priority areas to concentrate biodiversity conservation efforts. The use of remote sensing (RS) products in species distribution models and biodiversity monitoring (Radeloff et al., 2019), particularly the Dynamic Habitat Index (DHI), is therefore crucial to effectively deliver accurate and timely information on habitats and species distributions; guiding conservation planning and decision-making, and ultimately supporting the protection and conservation of biodiversity (Pettorelli et al., 2014; Préau et al., 2022a; Venter et al., 2016).

Integrating predictors derived from LULC classification into local-scale SDMs is considered a traditional approach, whilst DHI integration represents an innovative methodological step (Arenas-Castro et al., 2018; Coops and Wulder, 2019; Pettorelli et al., 2016; Randin et al., 2020; Regos et al., 2022). It will be argued that the incorporation of DHI into SDM models remains underutilised in the scientific community. While some studies combine LULC and DHI inputs into SDMs (Michaud et al., 2014; Zhi et al., 2022), scientific literature has not extensively compared the two approaches to explore their nuances. Comparative analysis may therefore improve our understanding of the advantages and disadvantages presented by each approach and elucidate their respective impact on species distribution modelling outcomes.

In this study, SDMs using DHI as predictors were evaluated to quantify their predictive performance. To this end, two distinct SDM approaches, commonly used in scientific literature, were empirically compared. The first modelling approach used DHIs derived from RS data; the second relied on landscape metrics derived from LULC classifications. Prediction scores were analysed before and after their transformation into binary maps (binarisation), used to represent predicted ecological niches, in order to evaluate the effectiveness of their habitat identification predictions. The LULC approach relies on common landscape metrics derived from LULC classifications, with Euclidean distance used as a measure of spatial proximity to specific land-cover types. In contrast, the RS approach uses DHI, continuous variables directly obtained from Sentinel-2 RS data, without considering distance. It is important to note that, although LULC data are also derived from RS, they undergo several transformations to produce categorical outputs, most notably via supervised classification. We expect our results to indicate that variables directly computed from RS products (namely DHIs) can be used to improve the accuracy SDM predictions. Since species distribution modelling is an operational tool, DHI integration is expected to improve the predictive accuracy of species dispersal models, providing tools for habitat quality monitoring and actionable information to support conservation and habitat management decisions at a regional scale.

## 2. Material and Methods

### 2.1 Study area

The study area encompassed the administrative region of Île-de-France (IDF) (12,000 km²). This region, dominated by intensive agriculture and urban areas, is the most densely populated and built-up region in France (Figure 1). It is relatively homogenous in terms of bioclimatic variables, with a temperate climate and very fertile agricultural soils. Île-de-France is located at the centre of the Paris Basin, a sedimentary basin with a relatively flat relief, and is irrigated by the River Seine, whose main tributaries converge in this region.

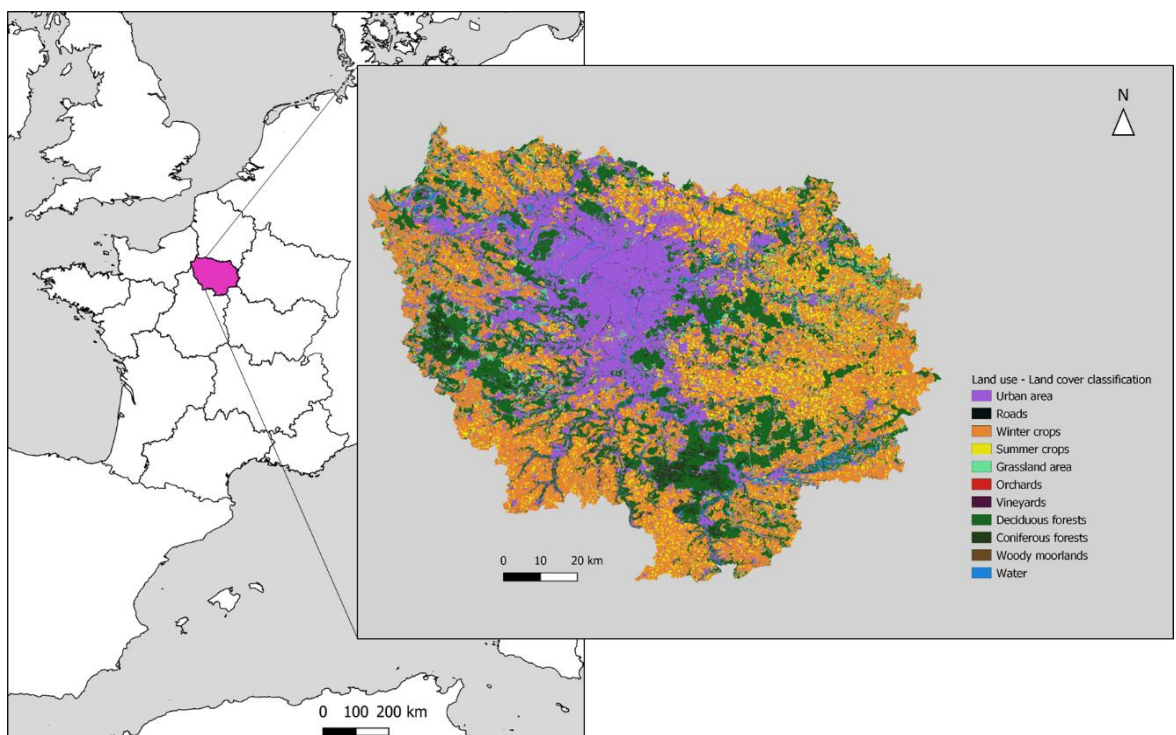

**Figure 1.** Study site location: the French region Île-de-France (2018 land-use/land-cover classification from CES OSO-CESBIO (CESBIO, 2020).

### 2.2 Data

**Species data –** We compiled data for eleven species for the period 2015-2020 from the Cettia ÎDF naturalist citizen database ("GeoNat'îdF," 2020), including four bird species (*Athene noctua*, *Anthus pratensis*, *Pyrrhula pyrrhula, Sylvia curruca*), five amphibians (*Bufo bufo*, *Hyla arborea*, *Ichthyosaura alpestris*, *Lissotriton vulgaris*, *Triturus cristatus*), and two mammals (*Eptesicus serotinus*, *Meles meles*). The minimum accuracy of geolocation is estimated at 50 metres. Details on species and the number of occurrences are available in Table S1.

**Environmental data –** We compared two different sources of input predictors for the SDMs. To that end, we compiled two sets of predictor variables, a first set based on classified land-cover data (LULC predictors) and a second set based on remote sensing data (RS predictors). LULC and RS data were derived from multitemporal Sentinel-2 data level 2A optical image series from 2018, with HSR of 10 m. Level 2A refers to data that have been radiometrically corrected for atmospheric effects and are accompanied by a cloud mask generated using the MAJA processor (Baetens et al., 2019). Sentinel-2 HSR images (Drusch et al., 2012) are freely available from the Theia distribution platform (https://catalogue.theia-land.fr). The annual mean, maximum and minimum DHIs values were computed for 18 time series of vegetation, soil, and water radiometric indices using Orfeo ToolBox v7.2.0. (Grizonnet et al., 2017). Predictor variable details are available in Table S2. The LULC data were derived from a supervised classification of the Sentinel-2 time series using a random forest classifier (Inglada et al., 2017). Regarding LULC, some of the original 23 land-use classes that were too detailed for this topic have been aggregated, such as urban areas or crop classes. We then computed predictor variables describing the area and distance to the nearest class for all eleven LULC classes, using FRAGSTATS v4.2 (McGarigal et al., 2022). The variables describing distance to a LULC class represent the Euclidean distance to the nearest pixel of that class. The LULC and RS predictor variables were upscaled to a 50 m spatial resolution for consistency with the occurrence data.

## 2.3. Species distribution models

Ensemble modelling was performed using version 3.5–3 of the *biomod2* R package (Thuiller et al., 2012).

**Selection of variables -** One variable from each group of collinear variables was selected based on Pearson correlation (r≥0.7) using the removeCollinearity function from the R package *virtualspecies* (Leroy et al. 2015). For the RS variable set, the same 10 variables were selected for all species (Meynard et al., 2019). For the LULC variable set, the relative importance of each variable was evaluated for each species using the *biomod2* R package. The relative importance of the variables, calculated independently of the chosen algorithms for modelling, corresponds to the Pearson coefficient between the initial predictions of the model and the predictions obtained after randomly permutating the evaluated variable (Hawkins et al., 2017). Ten permutations were performed and the ten most important variables were selected for inclusion in the models (Brun et al., 2020). See Figure 2 for variables retained per species.

**Spatial Sorting Bias correction -** To reduce sampling bias arising from multiple observations in the pixel, presence points were spatially rarefied to match the resolution of the environmental and species data (50 x 50 m) and to avoid point clustering, retaining only one randomly selected occurrence per pixel (Boria et al., 2014; Dubos et al., 2022; Sillero and Barbosa, 2021). The resulting number of occurrences at this spatial resolution is provided in Table S1.

**Pseudo-absence selection –** Each model was run with five sets of pseudo-absences, (PAs) randomly selected within the background. For all models, the number of PAs matched the number of presence records for each species (Barbet-Massin et al., 2012; Liu et al., 2019). Presences and PAs were assigned equal weight.

**Algorithms used and model runs -** The models were run with nine algorithms available in *biomod2*, namely: Generalised Additive Model (GAM), Generalised Linear Model (GLM), Multivariate Adaptive Regression Splines (MARS), Artificial Neural Networks (ANN), Flexible Discriminant Analysis (FDA),

Classification Tree Analysis (CTA), Generalised Boosting Models (GBM), Random Forest (RF), and Maximum Entropy (MAXENT). Three calibrations were performed per model.

**Ensemble Model building -** Habitat suitability maps for each species were generated using an Ensemble Model, calculated as the average of the habitat suitability scores from models exhibiting an area under the ROC curve (AUC-ROC) of ≥ 0.7 (Hanley and McNeil, 1982).

**Model performance assessment –** The predictive accuracy of SDMs was assessed using two indices based on cross-validation. The observation dataset was split into 70% for training and 30% for validation and assessed using two evaluation metrics: The Continuous Boyce Index (CBI) and the calibrated area under the ROC curve (cAUC). CBI is a reliability metric adapted to presence-only models that measures the accuracy of occurrence predictions relative to a random distribution of occurrences along the prediction gradient (Boyce et al., 2002; Hirzel et al., 2006). The cAUC is a discrimination metric for presence-absence models that distinguishes between presence and absence sites while minimising spatial sorting bias. This metric is calibrated against a geographic null model which to account for the effect of spatial sorting bias (Hijmans, 2012).

## 2.4. Overlap assessment between predicted niches from LULC and RS SDMs

**Binarisation of habitat suitability score maps -** To assess the spatial overlap between predicted niches from ensemble SDMs using LULC and RS predictors (LULC SDM and RS SDM respectively), habitat suitability maps were binarised into occupancy (presence-absence) maps, representing the predicted niches for ecological interpretation. The 10th percentile of occurrence suitability scores was used as a binary threshold (Gantchoff et al., 2022; Préau et al., 2022a). Values above the 10$^{th}$ percentile are suitable niches.

**Quantification of the overlap degree between predicted niches from RS and LULC SDMs –** To compare the overlap of predicted niches, three overlap indices widely used in the literature were calculated using the R package *sdmTools* (Warren et al. 2021): Schoener's *D* (Schoener 1968) (Equation 1), Warren's *I* (Warren et al., 2008) (Equation 2) and Spearman's rank (ρ; Phillips et al., 2009). We calculated these indices prior to the binarisation of suitability score maps into occupancy maps.

$$D(p_{RS}, p_{LU}) = 1 - \frac{\sum_i |p_{RS,i} - p_{LU,i}|}{2} \quad \textbf{(Equation 1)}$$

$$I(p_{RS}, p_{LU}) = 1 - \frac{\sqrt{\sum_i (p_{RS,i} - p_{LU,i})^2}}{2} \quad \textbf{(Equation 2)}$$

Where $p_{RS,i}$ and $p_{LU,i}$ are the normalised suitability scores ($\sum_i p_i = 1$) for the prediction $RS$ from RS SDM and the prediction $LU$ from LULC SDM in the grid cell $i$, respectively. For the Warren's *I*, the square root corresponds to the Hellinger distance.

**Comparison of predicted niches from RS and LULC SDMs -** The similarity between predicted niches from the LULC and RS SDMs was assessed by analysing shared information using global indicators based on the overlap, RS off-overlap, and LULC off-overlap parts according to the binary threshold. A quantitative comparison was conducted to evaluate the potential effect of the binary threshold, exploring: (i) overlap rates in LULC- and RS-predicted niches (Equations 3 and 4) across threshold variations; (ii) the overestimation of LULC and RS SDMs (Equation 5); and (iii) the distribution of suitability scores inside and outside binary threshold (10$^{th}$ percentile) predicted niches, according to the importance of predictor variables. $O_{LU}$ and $O_{RS}$, the overlap rates in LULC- and RS-predicted niches, respectively, are defined as the proportion of species presences in one niche that also occur in the other, calculated using Equations 3 and 4. $R$ is the ratio of off-overlap rates between $RS-$ and $LULC-$predicted niches, calculated using Equation 5.

$$O_{LU} = \frac{\sum_i (b_{LU,i} \times b_{RS,i})}{\sum_i b_{LU,i}} \quad \textbf{(Equation 3)}$$

$$O_{RS} = \frac{\sum_i (b_{LU,i} \times b_{RS,i})}{\sum_i b_{RS,i}} \quad \textbf{(Equation 4)}$$

$$R = \frac{\sum_i b_{RS,i} - \sum_i (b_{LU,i} \times b_{RS,i})}{\sum_i b_{LU,i} - \sum_i (b_{LU,i} \times b_{RS,i})} \quad \textbf{(Equation 5)}$$

Where $b_{LU,i}$ and $b_{RS,i}$ represent species presence in cell $i$ of the overlap grid between the LULC- and RS-predicted niches: $b_{x,i}$ ($x \in \{LU, RS\}$) corresponds to $p_{x,i}$ after binarisation, where $b_{x,i} = 1$ if the species is predicted to occupy the niche by SDM $x$, and 0 otherwise. The sum of the products $b_{LU,i}$ and $b_{RS,i}$ over all cells $i$ gives the number of presences that overlap in both predicted niches.

## 3. Results

### 3.1 Assessment of SDMs according to LULC and RS predictors

**Selection of variables -** For all species, based on the variable importance values for the LULC SDMs—up to a maximum of 0.41—no landscape metric is sufficient as the sole predictor for the SDMs (Figure 2a). Distance and area variables are equally important to LULC SDMs predictions. The annual mean NDVI (DHI $_{NDVI\ mean}$) is the most important variable in the RS SDMs for all species, with importance values ranging from 0.25 to 0.98. The DHI $_{NDVI\ mean}$ may suffice as the sole predictor in some SDMs, such as for the species *Bufo bufo, Ichthyosaura alpestris, Triturus cristatus, Lissotriton vulgaris* and *Pyrrhula pyrrhula.* Notably, the annual maximum NDVI is the only sufficient variable for the SDM of *Athene noctua* (Figure 2b).

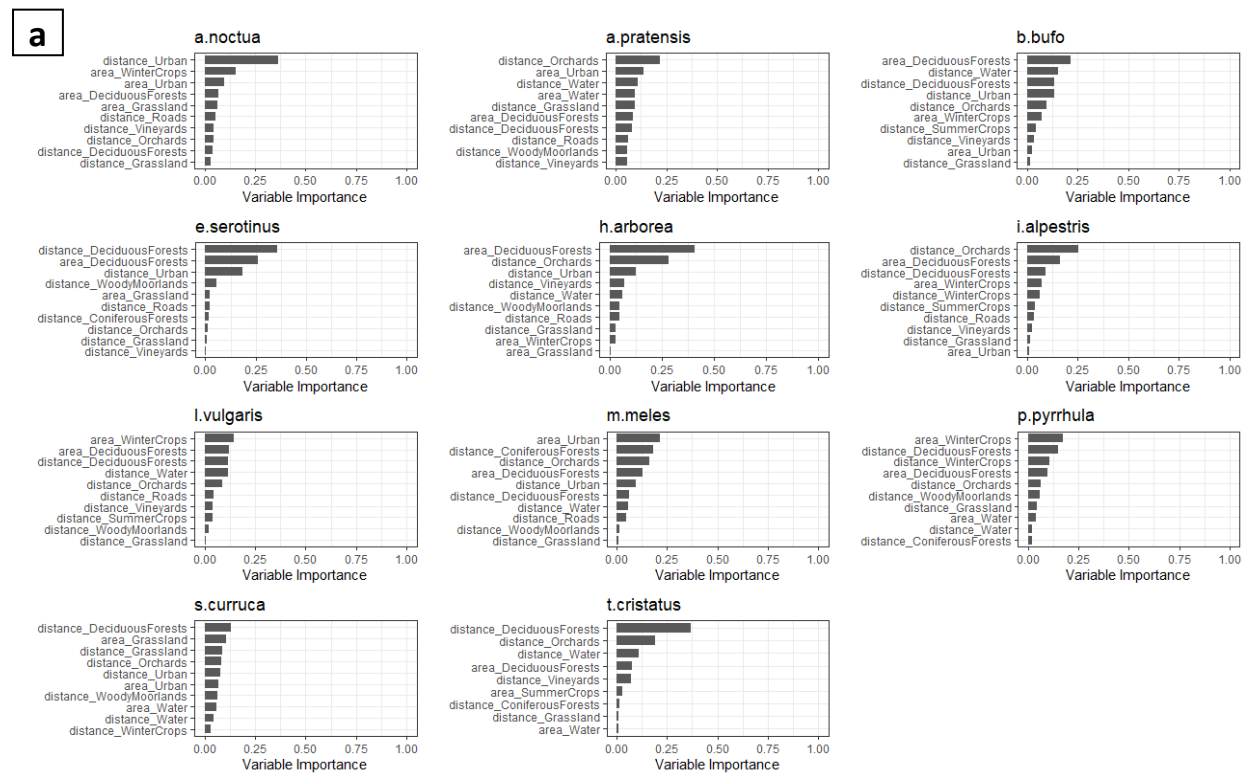

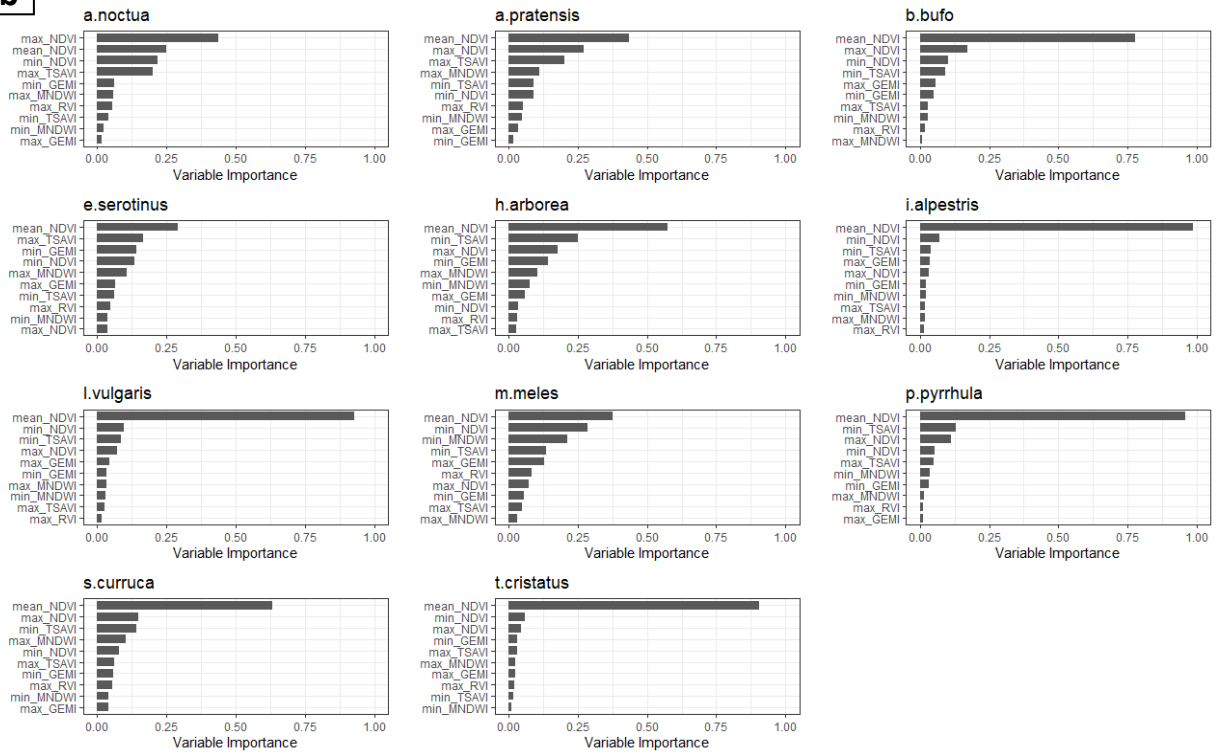

**Figure 2.** Variable importance of the ten SDM predictors by species for (a) LULC SDMs and (b) RS SDMs.

**Model performance assessment –** The evaluated contribution of algorithms to ensemble models is shown in Figure S1. Ensemble model performance, assessed using CBI, demonstrated strong and equivalent predictive accuracy of occurrences (values close to 1) for both LULC and RS SDMs, except for one species, *Eptesicus serotinus* (Figure 3a). Analysis with cAUC indicated that LULC and RS predictions performed similarly, consistent with CBI results. However, a more detailed examination revealed that LULC SDM predictions consistently outperformed RS SDM predictions across all species (Figure 3b). cAUC values tended towards 0.5 (ranging from 0.27 to 0.48), reflecting low discrimination between presence and absence sites for all species, irrespective of SDM type, but particularly for RS SDMs.

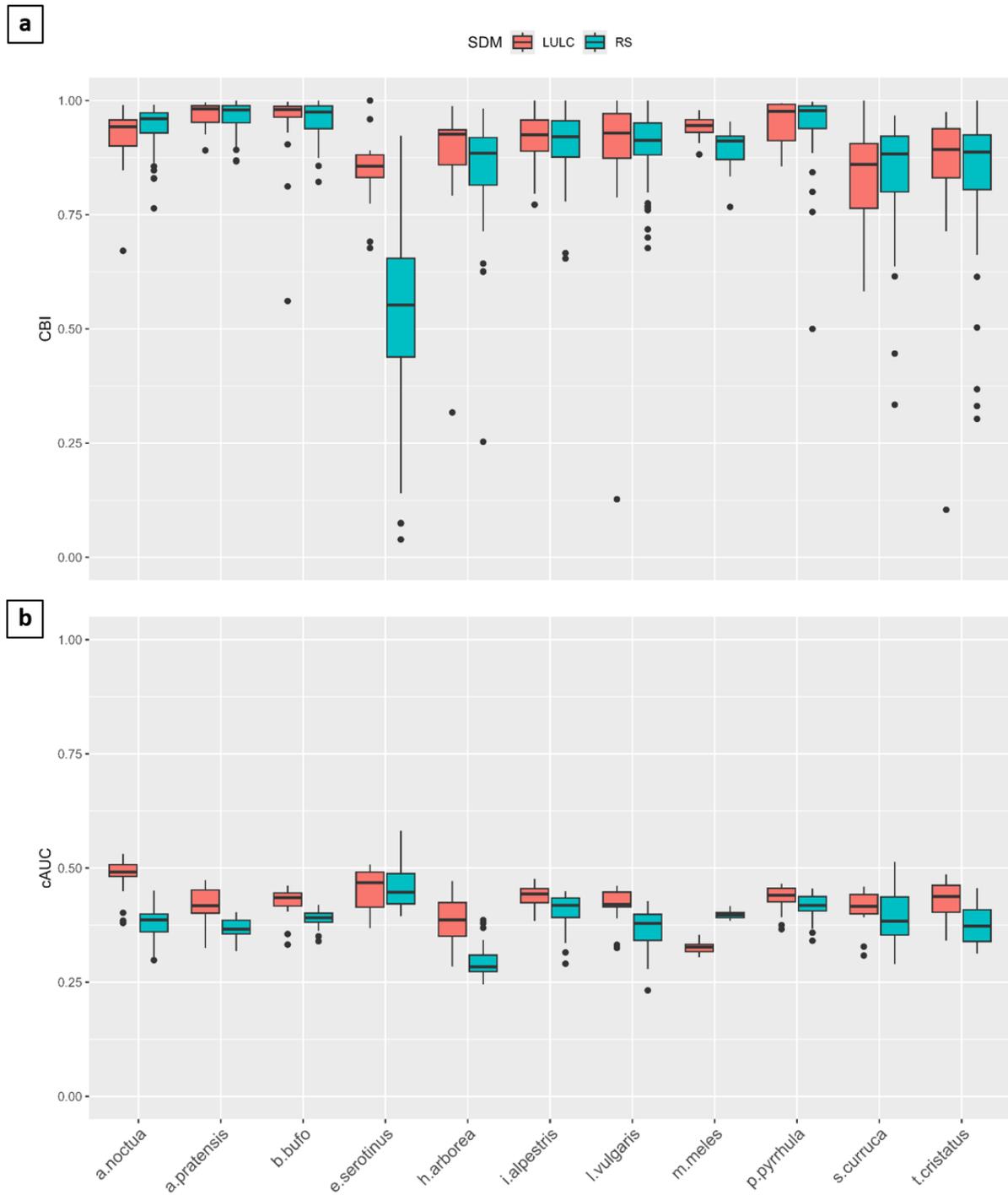

**Figure 3.** Assessment of LULC and RS ensemble SDMs by species using (a) Continuous Boyce Index (CBI) and (b) calibrated area under the ROC curve (cAUC).

## 3.2 Comparison of LULC and RS SDM results before and after binarisation: Overlap assessment

**Quantification of prediction overlap between RS and LULC SDMs –** The overlap between the LULC and RS suitability score maps (i.e. before the binarisation) is consistent with the similarity in prediction accuracy indicated by the Continuous Boyce Index, showing values close to 1 for *D* and *I* indices (min D = 0.72, min I = 0.93) but a more moderate ρ index (min ρ = 0.46) (Table 1).

| species | Schoener D | Warren I | Spearman ρ |
|---|---|---|---|
| *Athene noctua* | 0.827 | 0.972 | 0.757 |
| *Anthus pratensis* | 0.742 | 0.947 | 0.461 |
| *Bufo bufo* | 0.761 | 0.948 | 0.77 |
| *Eptesicus serotinus* | 0.781 | 0.949 | 0.702 |
| *Hyla arborea* | 0.799 | 0.961 | 0.745 |
| *Ichthyosaura alpestris* | 0.803 | 0.962 | 0.48 |
| *Lissotriton vulgaris* | 0.808 | 0.968 | 0.498 |
| *Meles meles* | 0.793 | 0.955 | 0.728 |
| *Pyrrhula pyrrhula* | 0.726 | 0.941 | 0.523 |
| *Sylvia curruca* | 0.722 | 0.934 | 0.537 |
| *Triturus cristatus* | 0.79 | 0.961 | 0.704 |

**Table 1.** Global indices used to evaluate the overlap between LULC and RS presence predictions for the eleven studied species before binarisation to 10$^{th}$ percentile threshold: Schoener's *D*, Warren's *I* and Spearman's rank (ρ).

**Comparison of predicted niches -** After binarisation, the metrics $O_{LU}$, $O_{RS}$ and $R$ (Table 2) revealed a greater extent of the predicted niche from the RS SDM compared to the LULC SDM, which can be observed spatially in the superposition of predicted niches in Figure 4. The $R$ metric showed that the LULC off-overlap was negligible compared to the RS off-overlap for all species, which on average was 2.61 ± 1.12 higher. The LULC off-overlap is characterised by pixel clusters surrounding areas of overlap, whereas the RS off-overlap consists of by numerous scattered, isolated pixels (Figure 4). The values of the $O_{LU}$ and $O_{RS}$ metrics support this observation, with predicted niches showing an average overlap of 73 ± 8% of the LULC-predicted niches (median 71% and 1$^{rst}$ quartile 68%), and 60 ± 15% of the RS-predicted niches (median 66% and 1$^{rst}$ quartile 45%).

| Species | Class of distance variables | Importance of distance variables | $O_{LU}$ | $O_{RS}$ | $R$ |
|---|---|---|---|---|---|
| *Athene noctua* | Urban | 0.36 | 0.71 | 0.4 | 3.68 |
| *Anthus pratensis* | Orchards | 0.22 | 0.67 | 0.58 | 1.51 |
|  | Water | 0.11 |  |  |  |
| *Bufo bufo* | Water | 0.15 | 0.85 | 0.68 | 2.69 |
|  | Urban | 0.14 |  |  |  |
| *Eptesicus serotinus* | Deciduous Forests | 0.35 | 0.78 | 0.67 | 1.74 |

| Species | Class of distance variables | Importance of distance variables | $O_{LU}$ | $O_{RS}$ | $R$ |
|---|---|---|---|---|---|
| Hyla arborea | Urban | 0.18 | 0.7 | 0.41 | 3.36 |
| | Orchards | 0.28 | | | |
| | Urban | 0.13 | | | |
| Ichthyosaura alpestris | Orchards | 0.25 | 0.83 | 0.48 | 5.23 |
| | Deciduous Forests | 0.09 | | | |
| Lissotriton vulgaris | Deciduous Forests | 0.12 | 0.77 | 0.62 | 2.01 |
| Meles meles | Deciduous Forests | 0.18 | 0.74 | 0.68 | 1.38 |
| | Orchards | 0.16 | | | |
| Pyrrhula pyrrhula | Deciduous Forests | 0.15 | 0.83 | 0.66 | 2.62 |
| | Winter Crops | 0.1 | | | |
| Sylvia curruca | Deciduous Forests | 0.13 | 0.64 | 0.43 | 2.37 |
| | Grassland | 0.09 | | | |
| Triturus cristatus | Deciduous Forests | 0.37 | 0.8 | 0.66 | 2.18 |
| | Orchards | 0.19 | | | |
| | Water | 0.11 | | | |

**Table 2**. Table summarising overlap metrics ($O_{LU}$, $O_{RS}$, $R$); *class of distance variables* (identifying up to three of the most important distance variable classes in the LULC SDMs); and *importance of distance variables* (average importance value for each model constituting the ensemble model) for each *species* studied.

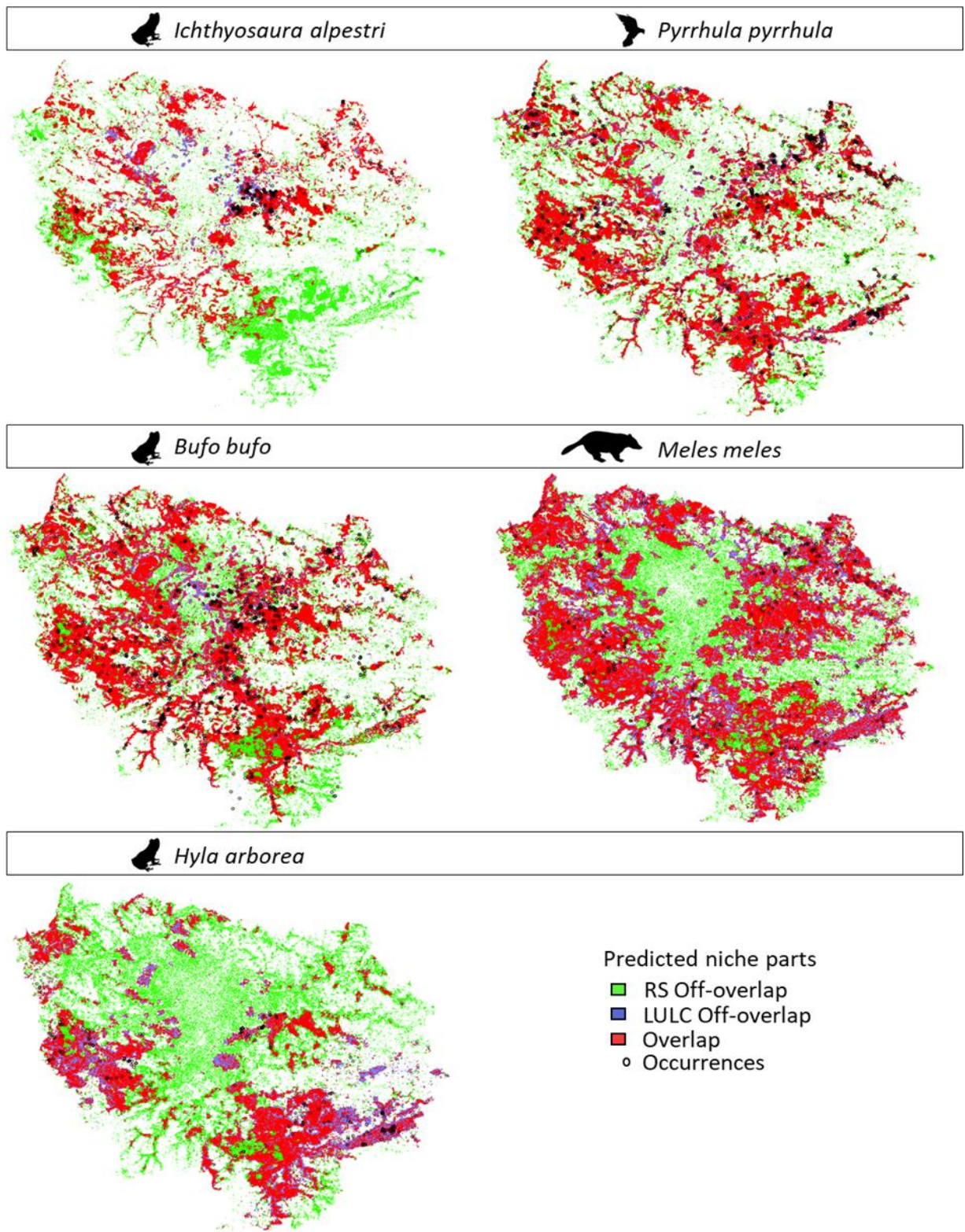

**Figure 4.** Overlap maps of LULC- and RS-predicted niches with distribution of occurrences for the species: *Ichthyosaura alpestris*, *Pyrrhula pyrrhula*, *Bufo bufo*, *Meles meles*, *Hyla arborea* (Overlap maps of other species in Figure S2).

**Overlap across binary thresholds –** Across all binary thresholds, the overlap projected onto the RS SDM predicted niche ($O_{RS}$,) was consistently lower than the overlap projected onto the LULC SDM predicted niche ($O_{LU}$) (Figure 5). This suggests that the RS approach predicts a broader niche than the LULC approach, irrespective of the binary threshold applied.

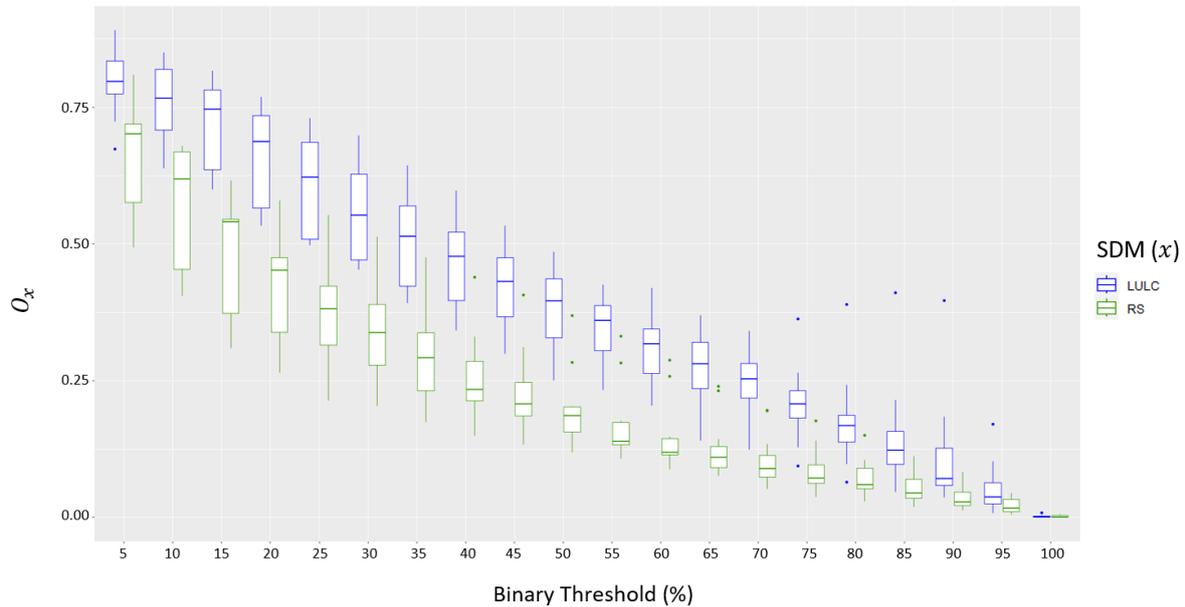

**Figure 5.** Synthesis of assessed Overlap metrics $O_{LU}$ and $O_{RS}$ ($O_x$) according to the binary threshold (%) across all studied species.

**Overlap quantitative assessment: distribution of important variables within predicted niches -** To better understand the differences between RS and LULC SDMs, we plotted response curves of suitability scores across the predicted niches according to the values of the most important variables in each SDM. The response curves for the most important variables in the SDMs revealed a distance effect, with suitability scores—transformed into presence/absence after binarisation — varying across different parts of the LULC-predicted niche according to distance to the LULC classes of interest. This means that the LULC-RS overlap and LULC off-overlap curves terminate abruptly at the same distance values, whereas the RS off-overlap and off-binary threshold curves extend toward higher distance values (Figure 6). Cross-referencing these results with Table 2 indicates that, for each of the eleven species studied, at least one distance-related variable ranks among the three most important variables in the LULC SDMs. In the LULC predicted niches, suitability are associated with the distance from the relevant LULC classes, whereas in the RS niches, continuous variables are not related to distance. Following binarisation, the further a score is from an LULC class, the more spatially constrained it becomes (Figure 4).

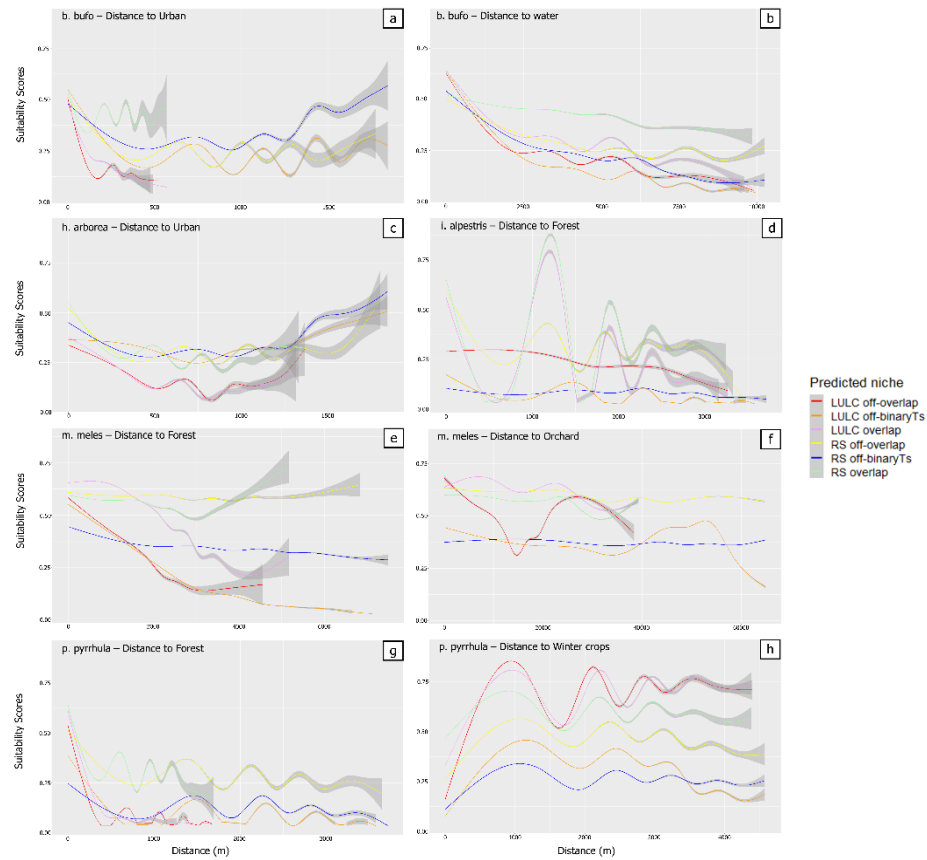

**Figure 6.** Response curves for the distance variable showing the effect of binarisation at 10[th] percentile binary threshold for the following species: *Ichthyosaura alpestris*, *Pyrrhula pyrrhula*, *Bufo bufo*, *Meles meles*, *Hyla arborea.* Off-binary Ts: parts of the LULC-/RS-predicted niches with suitability scores not selected by the binary threshold. Overlap: the portion of LULC-/RS-predicted niches selected by the binary threshold. RS overlap: RS suitability scores within the overlap. LULC overlap: LULC suitability scores within the overlap. Off-overlap: the portion of LULC-/RS-predicted niches outside the overlap but selected by the binary threshold.

**Relationship between the ratio of off-overlap parts of the predicted niches and the distribution of species occurrences –** Following binarisation, the ratio of off-overlap parts of the LULC- and RS-predicted niches (R) and the mean distance between species occurrences exhibited a generally decreasing relationship for most of the studied species. The exception to this pattern was observed in three species: *Athene noctua, Lissotriton vulgaris* and *Triturus cristatus* (Figure 7).

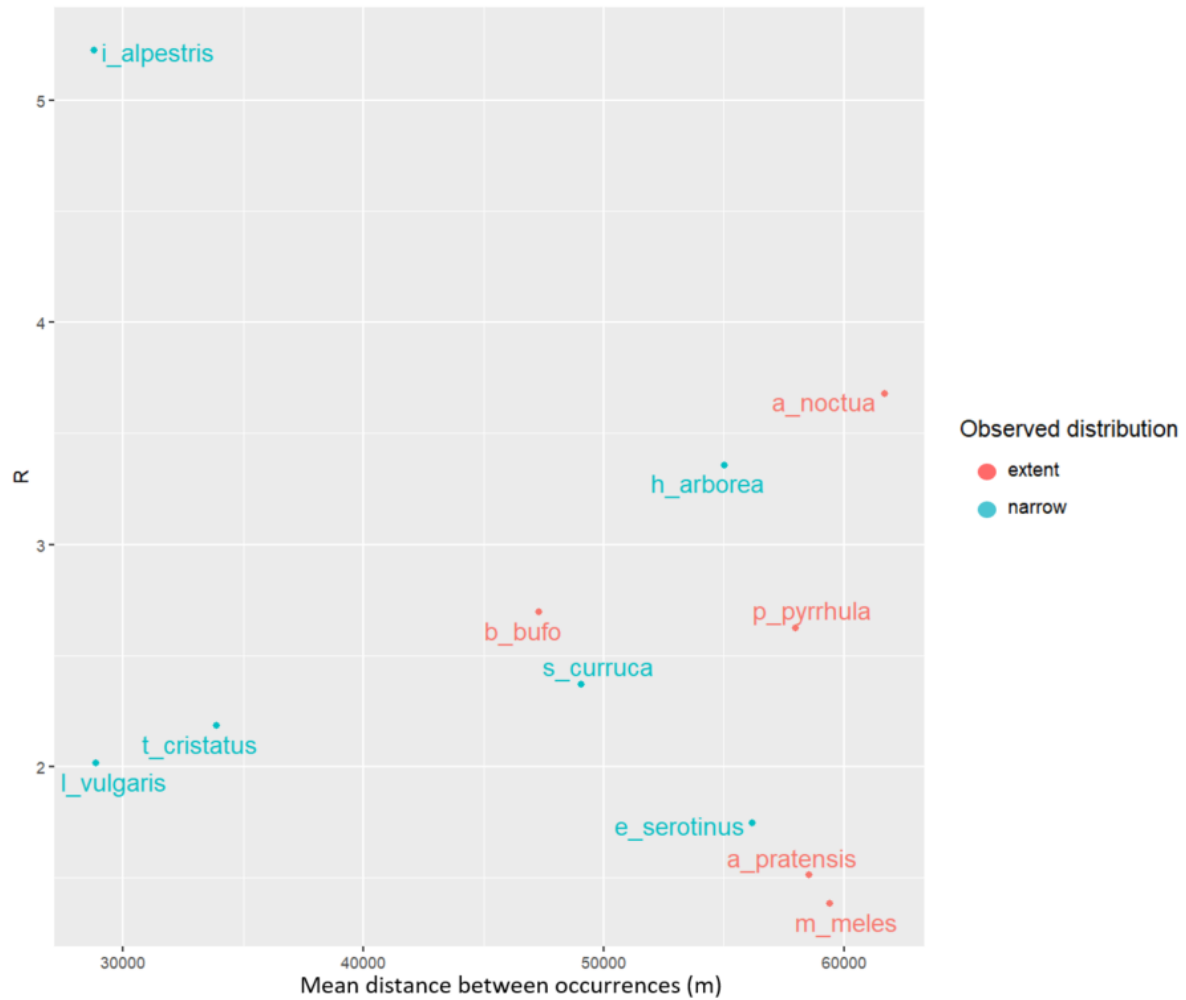

**Figure 7.** R, ratio of $RS-$ and $LULC-$predicted niche off-overlap, versus the mean distance between species occurrences (m) for the eleven studied species, showing observed distributions.

## 4. Discussion

### 4.1. DHI based on NDVI: the most important predictor in our RS approach

This study found that the NDVI-based annual mean DHI was the most significant indicator of habitat quality in the RS approach, with the potential to serve as the sole SDM predictor for most species. This finding is consistent with the widespread application of RS-derived NDVI in ecological studies (Nieto et al., 2015; Radeloff et al., 2019; Toszogyova and Storch, 2019). NDVI and related spectral vegetation indices are well-established proxies for above-ground net primary production (Chen and Coops, 2009) and ecosystem functioning (Alcaraz-Segura et al., 2017; Gonçalves et al., 2016; Pettorelli et al., 2014). Plant productivity, closely linked to NDVI, is recognised as a reliable measure of available food and territorial resources for animals (Fernández et al., 2016; Ramírez et al., 2017; Regos et al., 2022). Consequently, it has been used to effectively predict the distribution and richness of animal and plant species across spatial scales (e.g. Clauzel and Godet, 2020; Sheeren et al., 2014; Tarabon et al., 2019).

The annual mean DHI, derived from NDVI, proved to be the most relevant indicator of average annual plant productivity in this study. This is particularly advantageous in low-productivity environments such as our study region, Ile-de-France, which is characterised by heterogeneous landscapes dominated by an urban fabric interlaced with agricultural land. Unlike other vegetation products, NDVI-based DHI captures greater variability and avoids being underestimated as zero, thereby providing a more realistic measure of habitat quality (Radeloff et al., 2019). Moreover, its strong association with bird species richness highlights its potential for predicting avian biodiversity patterns across spatial scales, and evidence suggests that incorporating seasonal and interannual variation in NDVI-based DHI strengthens its predictive power at regional scales (Coops et al., 2009a).

Conversely, Bonthoux et al. (2018) question the use of DHI for studying bird communities in France, suggesting that models based on a single, carefully selected NDVI period might be more appropriate. At a fine scale of 4 km², they found that birds respond primarily to local landscape factors, and that single-date NDVI was more effective than NDVI-based DHI in discriminating habitat composition and heterogeneity, particularly in agricultural landscapes. These results corroborate prior studies, showing that NDVI-based measures of seasonal vegetation productivity outperform NDVI-based DHI as predictors (Hawkins, 2004; Hurlbert and Haskell, 2003). In addition, the study by Hobi et al. (2017) compared DHIs derived from various vegetation radiometric indices to explain breeding bird species richness and found that DHI based on NDVI or EVI was less effective than DHI calculated from fPAR, LAI, or gross primary productivity. This study did not incorporate these indices and, considering the findings of Bonthoux et al. (2018), emphasis placed on NDVI-based DHI should be interpreted with caution.

### 4.2. Strong and equivalent predictive accuracy of LULC and RS SDMs

It is widely recognised that RS-derived environmental predictors enhance SDM performance and predictive by supplying high-resolution data across broad spatial and temporal scales (Leitão and Santos, 2019; Randin et al., 2020; Regos et al., 2022). Requena-Mullor et al. (2014) reported that ecosystem functioning variables, based on interannual radiometric indices, outperformed LULC variables in predicting badger distribution at a local scale (835 km²) in the south-eastern Iberian Peninsula. However, the results of the current study did not support the hypothesis that a DHI-based RS approach, using annual time series of radiometric indices, improves SDM performance and accuracy at local or regional scales. Both RS and LULC approaches showed similar prediction accuracy, though RS-based SDMs discriminated less effectively between sites of presence and pseudo-absence. These findings corroborate Arenas-Castro et al. (2018), who evaluated LULC and RS approaches using radiometrically detected interannual time series as proxies for ecosystem dynamics. In their study, RS indices performed similarly to the climate-LULC combination, suggesting that they approximate several environmental factors. Arguably, the incorporation of HSR climate data, LULC classifications recently derived from RS, edaphic, and topographic data could further improve model performance. It is important to note that RS data are not always suitable for all species, as acknowledged by Leitão and Santos (2019) regarding the study of *Eptesicus serotinus*.

### 4.3. Comparison of ecological niche predictions derived from LULC and RS data

The results indicate a strong overlap between the habitat suitability score maps produced by the LULC and RS approaches for all the species studied. However, once these maps are binarised into predicted ecological niches, the overlap between the niches derived from these two approaches diminishes markedly.

**Analysis of off-overlap patches** – In the predicted niches after binarisation, the LULC off-overlap patches appear to connect overlap patches, contributing to a more coherent spatial structure while remaining concentrated around these zones of shared information. In contrast, RS off-overlap patches are often more dispersed, composed of isolated pixels, and sometimes cover a large portion of the study area (Figure 4). The differences in predicted niche structures can be attributed to the nature of the input data. LULC data, being categorised into discrete classes, generally produce spatially smoothed predictions with large, homogeneous patches. Conversely, RS data, characterised by their radiometric continuity, are not constrained by classes and exhibit greater variability in pixel distribution. This often results in dispersed, isolated pixels that can extend well beyond the main habitats forming the overlap. Such dispersion may lead to an overestimation of potential presence areas, interpreted as false positives by the AUC evaluation metric (Figure 3b). However, these areas could also contain elements of ecological connectivity not captured by observations due to sampling bias. Nevertheless, the dispersion of isolated pixels observed in the RS predictions likely reflects model overestimation rather than valid detection of biological corridors. These results are in line with several previous studies, where RS-based models tended to predict more fragmented areas than models based on LULC predictors (Arenas-Castro et al., 2018; Gonçalves et al., 2016).

By incorporating LULC classes and their spatial configuration, SDMs may capture biological corridors by identifying off-overlap areas that connect core habitat patches. Unlike RS data, LULC data provide directly interpretable information on the immediate landscape and facilitate an understanding of the relationships between landscape composition and species distribution with respect to biological corridors, ecological connectivity, and habitat fragmentation (Zeferino et al., 2020). We can therefore hypothesise that differing radiometric conditions may prevent RS-based SDMs from correctly identifying biological corridors, for instance, amphibians crossing wetlands encounter radiometric conditions markedly different from their typical habitats. However, a reliable detection of biological corridors requires a targeted approach that integrates data on ecological connectivity and species movements.

**Distance effect** – Distance variables consistently rank among the three most significant variables influencing LULC niche predictions across all species studied (Table 2). While LULC-predicted niches are spatially constrained, RS-predicted niches generally follow a trend whereby narrower species distributions—approximated by the average distance between occurrences—correspond to larger niche predictions (Figure 7). However, some species deviate from this trend, indicating that the spatial constraint acting on LULC-predicted niches depends not only on the overall distribution of occurrences but also on the extent and distribution of those occurrences within the relevant LULC classes. Take *Lissotriton vulgaris*, for example, a species with a narrow distribution, approximated by the average distance between occurrences. Whilst the RS- and LULC- predicted niches for *Lissotriton vulgaris* resemble those of other species, with a relatively low R (Figure 7), 70% of occurrences for this species are in forest patches (Table S1). In these patches, distance from this class is the third most important variable (Figure 2a). Thus, despite the relatively small average distance between occurrences in a main forest patch, numerous isolated occurrences exist in all other forest patches of the ÎDF (Figure S2). Consequently, the distribution of occurrences across the entire forest area limits the spatial constraint on the LULC-predicted niche, highlighting the influence of occurrence bias (Dubos et al., 2022; He et al., 2015).

The use of citizen occurrence databases, such as those used in this study, have inherent limitations, particularly spatial biases caused by uneven sampling effort. These biases can lead to an uneven distribution of occurrences across land-use classes, skewing predictions and misrepresenting actual species distributions. As a result, predicted niches may be underestimated in areas with sparse records density and an overestimation in areas with dense records (Beck et al., 2014; Bird et al., 2014; Botella et al., 2021; Dubos et al., 2022; Johnston et al., 2020). The binarisation process further revealed that LULC-based models are subject to a strong distance effect, with suitability scores decreasing as the distance occupied LULC classes increases. By contrast, RS-based models, which rely on continuous data, are not influenced by distance effects or geographical sampling biases. Nevertheless, the use of LULC distance metrics in SDMs remains common for both plant species (e.g. Linnell et al., 2017) and animal species ( e.g. Biscornet et al., 2021; Gantchoff et al., 2022; Li et al., 2020; Santika et al., 2014; Uriostegui-Velarde et al., 2018). It is therefore essential to account for the distance effect, which could be further explored using virtual species, and to recognise the significance of "area" variables, that describe landscape configuration and habitat distribution. Although they do not contribute to the spatial constraints in LULC-predicted niches, "area" variables are critical for understanding species-landscape relationships.

**Temporal resolution differences between RS- and LULC-based approaches –** Temporal resolution differences between RS- and LULC- based predictors may also contribute to variation in niche predictions. Remote sensing data, through the DHI, provide intra-annual measures of habitat productivity and variability, capturing seasonal habitat changes in structure, composition, and environmental conditions that static LULC data cannot (Alcaraz-Segura et al., 2017; Arenas-Castro et al., 2018; Regos et al., 2022). DHI also circumvents the challenges of regularly updating LULC classifications, making it particularly valuable for biodiversity monitoring, conservation, and restoration planning (Radeloff et al., 2019; Rose et al., 2015).

**LULC classification typology problem –** A further limitation lies in LULC classification typologies, which are not always well-suited to the ecological niches of target species. Imprecise land-use classification can lead to predictive errors (Bradley and Fleishman, 2008; Krishnaswamy et al., 2009). In this study, the OSO Land Cover classification (https://www.theia-land.fr/en/product/land-cover-map/) was tailored primarily for mammal and bird species. For instance, while the orchard class was scarcely represented across urban, forest, and agricultural landscapes in the Île-de-France (Figure 1), it was highly relevant for the European badger (*Meles meles*), which relies on orchard-forest mosaics for food and shelted (Requena-Mullor et al., 2014). Conversely, this classification was less suitable for amphibians; it includes only a single "water" class, excluding water bodies smaller than 10 m² that are critical for the reproductive and nesting needs of many amphibian species (Bartrons et al., 2024; Tornero et al., 2024). Nevertheless, with the exception of *Ichthyosaura alpestris*, all amphibians studied showed distributions closely associated with water, with distance to water ranking among the ten most important predictors (Figure 2). Despite these typological constraints, the selected landscape metrics and LULC classes proved effective, reflected in the strong performance of the SDMs (Figure 3).

## 5. Conclusion

In conclusion, given the inherent biases of each approach, this study does not allow us to definitively determine whether an RS -based or LULC -based approach is better suited for SDM integration, but highlights their respective strengths and limitations for modelling species distributions at regional scales. The RS approach tends to overestimate potential species habitats, whereas the LULC approach tends to underestimate them. These discrepancies likely stem from the modelling process, which relies exclusively on presence data and requires binarisation of predictions using a selected threshold. This step introduces potential errors, such as false presences or absences, resulting from

under- or overestimation of the niche (He et al., 2015). Despite these limitations, our results consistently show that the RS approach predicts a more expansive ecological niche than the LULC approach, irrespective of the threshold applied (Figure 6). In the context of conservation planning, it is therefore especially important to note that RS-based models generally predict broader ecological niches than LULC-based models. This tendency can be advantageous for exploring or identifying new areas of potential species occurrence. Taken together, RS and LULC approaches provide complementary insights that improve our understanding of biodiversity responses to environmental change and promote more effective habitat conservation and management.

Our study confirms that the effectiveness of RS and LULC approaches to habitat characterisation and species distribution predictions depends on a variety of factors (Arenas-Castro et al., 2018; Cord et al., 2014; Dubos et al., 2022; Gonçalves et al., 2016). These factors include: the type of species studied, the distribution and abundance of occurrences, the spatial scale of the study area, the type of LULC data used (landscape metrics, LULC classes), and the spatial and temporal resolution of the RS data used. It is important to take these factors into consideration when selecting the approach and data to be implemented in SDMs.

The comparison of RS and LULC approaches provides valuable insights into their respective contributions and highlights their complementarity; while RS data offer detailed insights into vegetation and land use changes, LULC data describe the spatial structure and configuration of habitats. As evidenced by numerous studies, the collinearity of predictors means that a combined approach can significantly enhance SDM performance (e.g. Bellón et al., 2022; Gantchoff et al., 2022; Kosicki et al., 2015), particularly when traditional environmental variables derived from LULC data are integrated with intra and interannual RS data (Arenas-Castro et al., 2018; Coops and Wulder, 2019; Michaud et al., 2014; Mod et al., 2016; Radeloff et al., 2019; Zhi et al., 2022). By integrating these predictors, environmental characteristics such as vegetation cover, habitat composition, and other landscape factors can be captured across various spatial and temporal scales. This comprehensive approach may offer a more accurate representation of ecological processes and the interactions between species and their environment (Le Roux et al., 2017; Préau et al., 2022b), improving our ability to predict the effects of environmental, structural, and habitat functional changes on biodiversity and guiding effective habitat conservation and management strategies. However, special attention is needed for narrow-range endemic species, which pose particular modelling challenges due to their limited occurrence data, fine-scale habitat requirements, and sensitivity to spatial resolution. Our results highlight the importance of integrating high-resolution and locally informed datasets, considering ensemble approaches that combine RS and LULC predictors, and explicitly accounting for uncertainty. We recommend that predictions for these species be used primarily as exploratory tools to guide field validation and conservation prioritization, rather than as definitive range estimates.

By combining RS and LULC data, conservation planners can identify critical areas where habitat loss or degradation is most likely to occur, enabling targeted interventions to preserve key ecosystems. For instance, RS data can detect early signs of habitat fragmentation which enables timely action to be taken to mitigate this threat. Meanwhile, LULC data can help planners understand the broader context, ensuring that conservation efforts effectively maintain ecological connectivity and support species movement across fragmented habitats. This integrated approach can inform the design of protected areas and ecological corridors, ensuring that these spaces not only cover areas of high biodiversity but also maintain the ecological functions necessary for species survival. By enhancing the predicted impacts of climate change on different habitats, adaptive management strategies can be developed to accommodate changes in species distribution and environmental conditions. In summary, incorporating both RS and LULC predictors into conservation planning efforts may enable dynamic and responsive strategies to be developed which address current threats and anticipate future challenges to ecosystems, ensuring more resilient conservation outcomes.

## CRediT authorship contribution statement

Maïri Souza Oliveira: Conceptualization, Methodology, Formal analysis, Investigation, Validation, Writing – original draft, Writing – review and editing.

Clémentine Préau: Conceptualization, Data curation, Methodology, Investigation, Writing – review.

Samuel Alleaume: Conceptualization, Methodology, Formal analysis, Investigation, Writing – review, Funding acquisition.

Maxime Lenormand: Conceptualization, Methodology, Investigation, Writing – review.

Sandra Luque: Conceptualization, Methodology, Investigation, Writing – review, Funding acquisition, Project administration, Supervision.

## Declaration of competing interest

The authors declare that they have no known competing financial interests or personal relationships that could have appeared to influence the work reported in this paper.

## Acknowledgments

We thank the THEIA CES Biodiversity variable at Centre National des Etudes Spatiales (CNES) for funding this study under the TOSCA project and for providing value-added data processed by CNES for the Theia data centre (www.theia-land.fr) using Copernicus products. The processing involved algorithms developed by THEIA's Scientific Expertise Centres. We also extend our gratitude to the organisations and observers contributing to the citizen data extracted from the regional tool Cettia ÎdF (https://www.arb-idf.fr/article/cettia-ile-de-france-fait-peau-neuve/).

## Data availability

All data used in this study are publicly available and have been cited in the main text and references.

# Supplementary material

**Table S1.** Exploration of the occurrence data of the studied species: "class" and "species" representing the species class and name, "abundance(res 50)" indicating the species abundance at a 50 m spatial resolution, "nb classes", "class code", "class name" and "class abundance (%)" corresponding to the number, codes, names and proportion of pixels of the LULC classes where species occurrences are located.

| class | species | abundance (res 50) | nb classes | class code | class name | class abundance (%) |
|---|---|---|---|---|---|---|
| Aves | Athene noctua | 419 | 5 | 1 | urban area | 53.46 |
| | | | | 5 | winter crop | 22.91 |
| | | | | 13 | grassland | 17.42 |
| | | | | 8 | summer crop | 3.34 |
| | | | | 16 | deciduous forest | 2.86 |
| | Anthus pratensis | 968 | 8 | 1 | urban area | 33.16 |
| | | | | 5 | winter crop | 24.38 |
| | | | | 23 | wetland | 16.12 |
| | | | | 13 | grassland | 14.05 |
| | | | | 8 | summer crop | 6.71 |
| | | | | 16 | deciduous forest | 4.55 |
| | | | | 17 | coniferous forest | 0.72 |
| | | | | 19 | woody heath | 0.31 |
| | Pyrrhula pyrrhula | 1129 | 8 | 16 | deciduous forest | 53.94 |
| | | | | 1 | urban area | 25.86 |
| | | | | 23 | wetland | 8.86 |
| | | | | 13 | grassland | 6.2 |
| | | | | 5 | winter crop | 2.48 |
| | | | | 17 | coniferous forest | 1.95 |
| | | | | 8 | summer crop | 0.53 |
| | | | | 19 | woody heath | 0.18 |
| | Sylvia curruca | 158 | 8 | 1 | urban area | 35.44 |
| | | | | 16 | deciduous forest | 21.52 |
| | | | | 13 | grassland | 21.52 |
| | | | | 23 | wetland | 10.76 |
| | | | | 5 | winter crop | 6.33 |
| | | | | 8 | summer crop | 1.9 |
| | | | | 19 | woody heath | 1.9 |
| | | | | 17 | coniferous forest | 0.63 |
| Mammalia | Eptesicus serotinus | 138 | 6 | 16 | deciduous forest | 47.83 |
| | | | | 1 | urban area | 35.51 |
| | | | | 13 | grassland | 6.52 |
| | | | | 23 | wetland | 4.35 |
| | | | | 5 | winter crop | 4.35 |
| | | | | 8 | summer crop | 1.45 |
| | Meles meles | 424 | 9 | 16 | deciduous forest | 46.23 |
| | | | | 5 | winter crop | 20.99 |
| | | | | 1 | urban area | 18.63 |
| | | | | 8 | summer crop | 5.42 |
| | | | | 13 | grassland | 4.95 |
| | | | | 4 | roads | 1.42 |
| | | | | 17 | coniferous forest | 1.18 |
| | | | | 19 | woody heath | 0.71 |
| | | | | 23 | wetland | 0.47 |

| class | species | abundance (res 50) | nb classes | class code | class name | class abundance (%) |
|---|---|---|---|---|---|---|
| Amphibia | Bufo bufo | 1213 | 9 | 16 | deciduous forest | 45.34 |
| | | | | 1 | urban area | 36.27 |
| | | | | 23 | wetland | 6.27 |
| | | | | 13 | grassland | 5.19 |
| | | | | 5 | winter crop | 4.29 |
| | | | | 17 | coniferous forest | 1.65 |
| | | | | 8 | summer crop | 0.74 |
| | | | | 19 | woody heath | 0.16 |
| | | | | 4 | roads | 0.08 |
| | Hyla arborea | 206 | 8 | 16 | deciduous forest | 35.92 |
| | | | | 1 | urban area | 24.76 |
| | | | | 13 | grassland | 15.53 |
| | | | | 23 | wetland | 10.19 |
| | | | | 5 | winter crop | 8.74 |
| | | | | 17 | coniferous forest | 1.94 |
| | | | | 8 | summer crop | 1.46 |
| | | | | 19 | woody heath | 1.46 |
| | Ichthyosaura alpestris | 297 | 6 | 16 | deciduous forest | 82.15 |
| | | | | 1 | urban area | 14.81 |
| | | | | 13 | grassland | 1.35 |
| | | | | 17 | coniferous forest | 1.01 |
| | | | | 5 | winter crop | 0.34 |
| | | | | 23 | wetland | 0.34 |
| | Lissotriton vulgaris | 294 | 7 | 16 | deciduous forest | 70.41 |
| | | | | 1 | urban area | 18.71 |
| | | | | 23 | wetland | 5.78 |
| | | | | 13 | grassland | 2.04 |
| | | | | 5 | winter crop | 2.04 |
| | | | | 17 | coniferous forest | 0.68 |
| | | | | 19 | woody heath | 0.34 |
| | Triturus cristatus | 216 | 8 | 16 | deciduous forest | 68.52 |
| | | | | 1 | urban area | 16.67 |
| | | | | 13 | grassland | 6.48 |
| | | | | 5 | winter crop | 4.17 |
| | | | | 23 | wetland | 1.85 |
| | | | | 8 | summer crop | 1.39 |
| | | | | 17 | coniferous forest | 0.46 |
| | | | | 19 | woody heath | 0.46 |

**Table S2.** Exploration of the predictor variables for both approaches, LULC and RS.

| Land Use - Land Cover (LULC) | | | Remote Sensing (RS) | | |
|---|---|---|---|---|---|
| Class | Code | Variables | Radiometric Index | Code | Variables |
| Urban area | 1 | | Brilliance Index | BI | |
| Roads | 4 | | Brilliance Index | BI2 | |
| Winter crops | 5 | | Color Index | CI | |
| Summer crops | 8 | | Global Environment Monitoring Index | GEMI | |
| Grassland area | 13 | area, distance from closest pixel of the same class | Infrared Percentage Vegetation Index | IPVI | |
| Orchards | 14 | | Index Surfaces Built | ISU | |
| Vineyards | 15 | | Modified Normalized Difference Water Index | MNDWI | |
| Deciduous forests | 16 | | Modified Soil Adjusted Vegetation Index | MSAVI | min DHI, max DHI, mean DHI |
| Coniferous forests | 17 | | Modified Soil Adjusted Vegetation Index | MSAVI2 | |
| Woody moorlands | 19 | | Normalized Difference Turbidity Index | NDTI | |
| Water | 23 | | Normalized Difference Vegetation Index | NDVI | |
| | | | Normalized Difference Water Index | NDWI | |
| | | | Normalized Difference Water Index | NDWI2 | |
| | | | Redness Index | RI | |
| | | | Ratio Vegetation Index | RVI | |
| | | | Soil Adjusted Vegetation Index | SAVI | |
| | | | Transformed NDVI | TNDVI | |
| | | | Transformed Soil Adjusted Vegetation Index | TSAVI | |

**Figure S1.** Performance evaluation of the nine modelling algorithms across all species based on AUC values for their inclusion in ensemble models (AUC ≥ 0.7) for (a) LULC and (b) RS approaches.

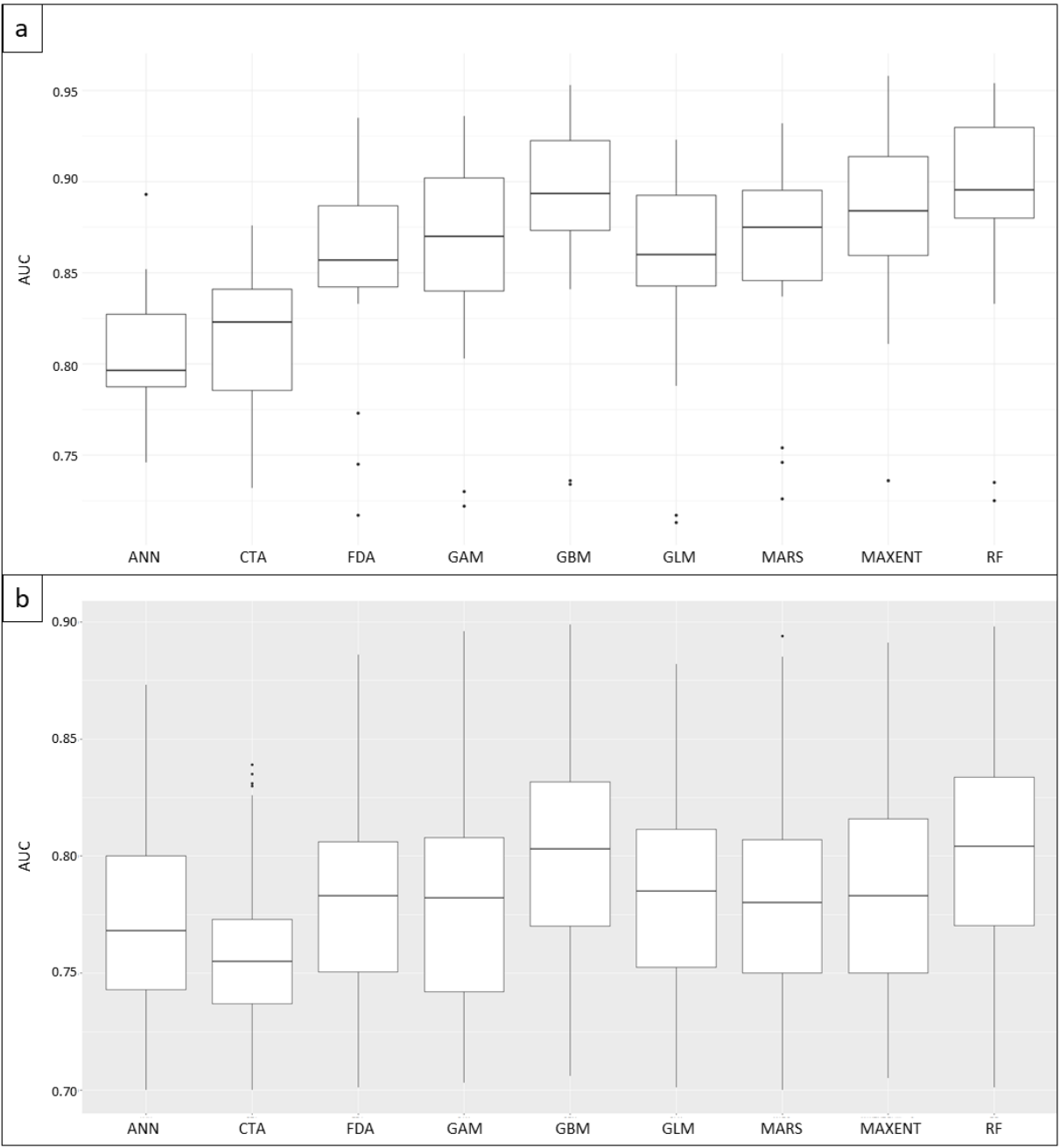

**Figure S2.** Overlap maps between LULC and RS predicted niches with distribution of occurrences for the species: *Lissotriton vulgaris, Athene noctua, Triturus cristatus, Eptesicus serotinus, Anthus pratensis, Sylvia curruca*.

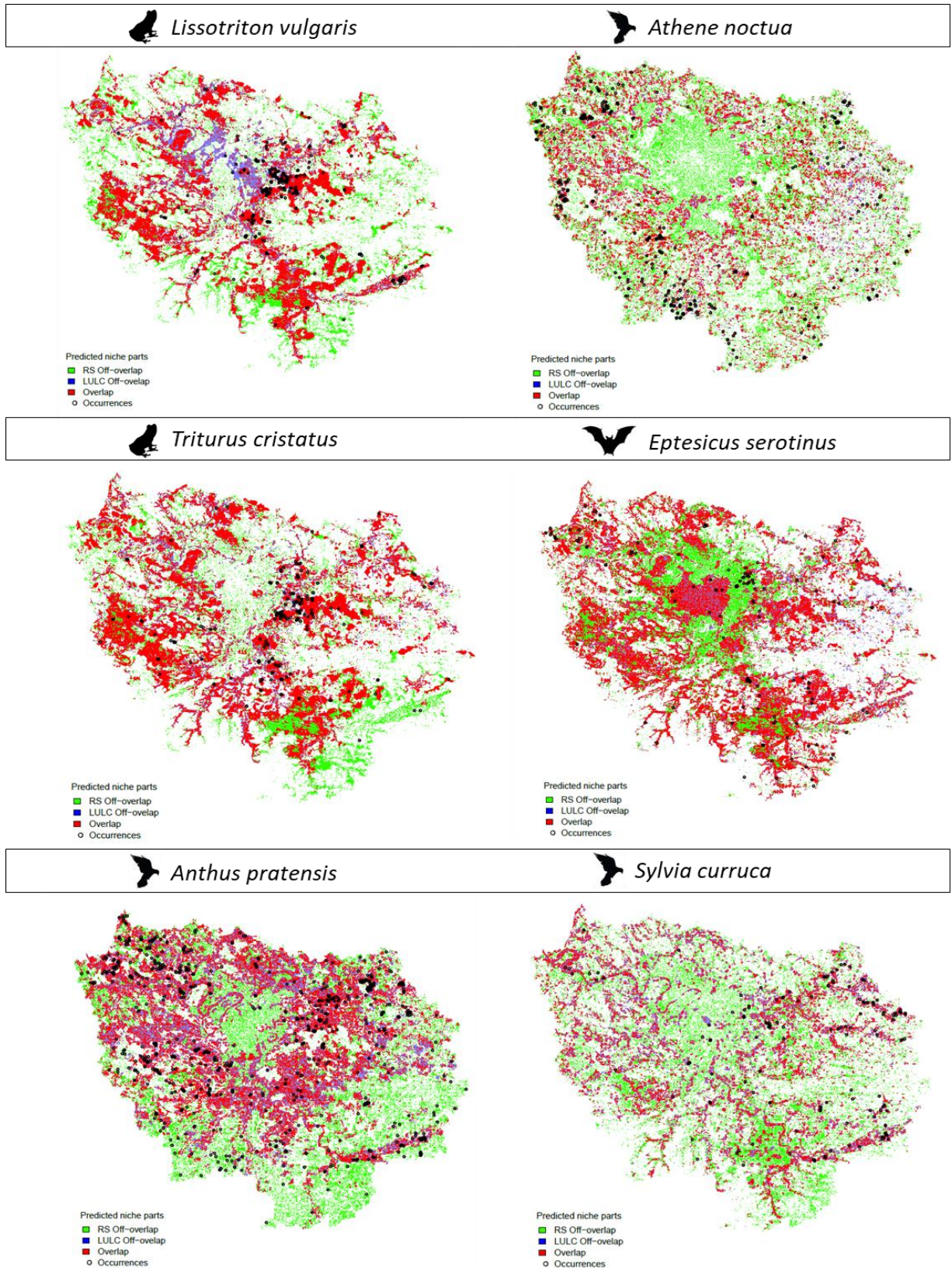